# Opinion mining using Double Channel CNN for Recommender System

1st Minoo Sayyadpour
*Dept. Math and Computer Science*
*Kharazmi University*
Tehran, Iran
std_minoosayyadpour@khu.ac.ir

2nd Ali Nazarizadeh
*Department of Computer Engineering*
*Central Tehran Branch, Islamic Azad University*
Tehran, Iran
computer.engineer.as@gmail.com

*Abstract*— Much unstructured data has been produced with the growth of the Internet and social media. A significant volume of textual data includes users' opinions about products in online stores and social media. By exploring and categorizing them, helpful information can be acquired, including customer satisfaction, user feedback about a particular event, predicting the sale of a specific product, and other similar cases. In this paper, we present an approach for sentiment analysis with a deep learning model and use it to recommend products. A two-channel convolutional neural network model has been used for opinion mining, which has five layers and extracts essential features from the data. We increased the number of comments by applying the SMOTE algorithm to the initial dataset and balanced the data. Then we proceed to cluster the aspects. We also assign a weight to each cluster using tensor decomposition algorithms that improve the recommender system's performance. Our proposed method has reached 91.6% accuracy, significantly improved compared to previous aspect-based approaches.

## I. Introduction

Recommender systems are inevitable today in our daily online journeys, from e-commerce to advertisement. In a very general way, recommender systems are algorithms aimed at suggesting relevant items to users. Recommender systems are critical in some industries as they can generate a tremendous amount of income when they are efficient or also be a way to stand out significantly from competitors. There are two significant paradigms of recommender systems: collaborative and content-based methods.

Collaborative methods for recommender systems are based on the past interactions recorded between users and items to produce new recommendations. These interactions are stored in the so-called "user-item interactions matrix". The main idea that rules these methods is that these past user-item interactions are sufficient to detect similar users and/or similar items and make predictions based on these estimated proximities.

The class of collaborative filtering algorithms is divided into two sub-categories: memory-based and model-based approaches. Memory-based approaches directly work with recorded interaction values, assuming no model, and are essentially based on the nearest neighbor's search. Model-based approaches assume an underlying "generative" model that explains the user-item interactions and tries to discover them to make new predictions.

Content-based approaches use additional information about users and/or items. This additional information can be, for example, the age, sex, job, or any other personal information for users, as well as the category or different aspects of items. Content-based methods aim to build a model based on the available "features" that explain the observed user-item interactions [1].

Among all of those techniques, collaborative filtering (CF) receives extensive attention and progress. This paper proposes an integrated opinion mining and CF method for enhancing the predictive performance of the Recommender Systems [2]. The model comprises two components: (1) weighted aspect-based opinion mining and (2) Recommendation generation. In the first component, after applying SMOTE algorithm to our dataset, we mainly designed a Double-channel Convolutional Neural Network that utilizes two different input layers, word embedding, and POS tag, for the aspect extraction task. We use an aspect-weighted rating estimation method based on the Tensor Factorization technique introduced in [2] to enhance the Recommender systems' predictive performance and evaluate it with and without SMOTE algorithm.

The most important contributions of this paper are listed below:

- For preprocessing: This article uses chat expansion to expand all the words used in chats and has more words as neural network input.

- For the aspect-sentiment classification task: This paper discusses the effect of combining a convolutional neural network (CNN) and using two pre-trained embeddings for aspect extraction and sending it to a weighting mechanism for clustering aspects. The first time these techniques are used together to accomplish this task.

- Presenting a rate prediction based on SVD decomposition and generating recommendations for a specific user.

- Highlighting future research and open issues in sentiment analysis.

The remainder of this paper is organized as follows: Section 2 describes the related articles and research in opinion mining and recommender systems, and Section 3 explains our proposed method. Section 4 evaluates our presented method and other selected approaches and compares accuracy, precision, and f1-score. Finally, Section 5 presents the conclusion and discusses challenges, open issues, and future trends.

## II. Backgrounds

This section summarizes studies that have applied methods used in sentiment analysis and recommender systems. There are papers which reviewed these approaches [3] or grouped them as ensemble approach [4]. For the past years, several approaches have been introduced for Aspect-based opinion mining on user text reviews [5] that are categorized into unsupervised methods such as rule-based [6] and topic modeling [7] and supervised approaches [8].

Several papers have recently used deep learning techniques for aspect-based opinion mining. [9] in 2014, a recurrent neural network (RNN) model was used for aspect extraction. One year later, [10] improved RNN by incorporating other linguistic features like part-of-speech (POS) tags and chunk information to guide the training and to learn a better model. [11] used a multi-layer CNN for aspect extraction. Another study with the CNN model is [12] that proposed two types of pre-trained embeddings for aspect extraction: general-purpose embeddings and domain-specific embeddings. Approaches for classifying the user sentiment polarity using Support Vector Machine (SVM) [13] and integrated neural network [14] in 2017 that combines CNN and Bi-LSTM for sentiment analysis. Lexicon-based approaches such as [15] use a statistical model for aspect ratings to discover the related topics. [16] propose a unified framework that identifies the product aspects and estimates the reviewer's weights. In this paper, we adopt the strategy used in [2,17] for Aspect-based opinion mining and generating recommendations and also use SMOTE algorithm [18] to balance data.

For recommend generation from review text, several works have been demonstrated. One of the earliest works to extract aspects as opinion targets and use them as features for collaborative filtering uses manually designed ontologies to generate free text [19]. This method is time-consuming and domain independent. To improve prediction performance, [20] combines latent rating dimensions with latent review topics such as those learned by LDA, and [21] combines content-based filtering with collaborative filtering by topic modeling. In 2014, [22] proposed a probabilistic model based on collaborative filtering and topic modeling to simultaneously model aspect ratings and user sentiments for improving the overall rating prediction. [23] presented the Explicit Factor Model (EFM) to produce recommendations and disrecommendations according to the specific product features for a particular user.

In recent years most papers have used deep learning methods for recommender systems. For example, CNN-based studies like [24] exploited two parallel neural networks coupled in the last layers that focus on learning user behaviors exploiting user's reviews, and learning item properties from the reviews written for the item. [25] developed that by introducing an additional latent layer representing the target user-target item pair.

Some studies used RNN instead for feature extraction. For example, [26] co-learns user and item information from ratings and customer reviews by optimizing matrix factorization and an attention-based GRU network. [27] presented an LSTM model to estimate user future listening behavior. [28] proposed a semi-supervised topic model to extract the product aspects and the associated sentiment lexicons from the textual user reviews with LSTM.

Auto Encoder approaches have also been used for recommender systems. [29] used stacked denoising autoencoders to perform feature extraction. Another study used a Model-based Collaborate Filtering Algorithm Based on a Stacked AutoEncoder to overcome the sparsity problem [30]. Also, [31] proposed a Personalized Recommendation method, which extends the item's feature representations with Knowledge Graph via a dual-autoencoder.

Another approach that matches our proposed method is the aspect-based approach. For instance, [32] presented jointly learned ranking models for individual aspects by modeling the dependencies between assigned ranks. [33] presented an entirely unsupervised approach for simultaneous aspect extraction and rating prediction, which does not require knowledge of the aspect-specific ratings or genres for inference. [34] introduced a novel probabilistic rating regression model to solve Latent Aspect Rating Analysis. This model was developed in [35] that does not require aspect keywords. [36] used the latent factor strategy for modeling the relationship between the aspects, user and product rating prediction, that extracts specific aspects of consumption of the items to enhance the user experience with those items further. [37] introduced the CNN model employing two types of pre-trained embeddings for aspect extraction: general-purpose embeddings and domain-specific embeddings. [38] presented an improved recommendation system with aspect-based sentiment analysis that replaces the attention sublayers with a simple, fast Fourier transform in the input embedding to model heterogeneous semantic relationships in a text.

Our paper is related to a method in [39] composed of three components, an opinion mining component, an aspect weighting computing component, and a rating inference component. Our proposed method uses an aspect extraction method based on a deep learning technique. Instead of directly using the extracted opinions from the text review, we also use a SMOTE method to balance data before estimating the aspect-weighted ratings.

III. PROPOSED APPROACH

This section presents our proposed recommender system based on an opinion mining approach. Our proposed algorithm includes two main parts: the aspect-based opinion mining unit, which is designed to extract aspects from users' opinions and assign weight to these aspects for each user. The second recommends a generation unit, which results in overall rating prediction using collaborative filtering. This unit increases the accuracy of recommendations. First, the SMOTE algorithm is applied to balance our data, and a Deep Neural Network extracts aspects. These steps are demonstrated in Fig. 1.

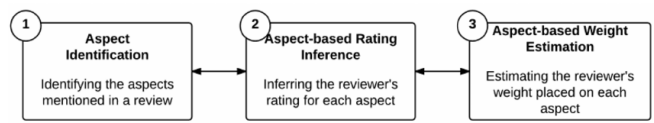

Fig. 1. Aspect Weight estimation Process [16].

After applying collaborative filtering, the second part determines five suggested products for the target user.

Analyzing and predicting ratings improves customer service, so the method proposed in this paper is essential.

*A. Preprocessing data*

Firstly, we added a column to categorize positive and negative comments and made a two-class dataset. In this column, rates below three are classified as negative and above three as positive.

We use Beautiful Soup [40] to remove HTML and XML tags, a Python package for parsing HTML and XML documents (including having malformed markup, i.e., non-closed tags, so named after tag soup). After that, we used pycontractions [41] for expanding and creating common English contractions in text. This function is helpful for dimensionality reduction by normalizing the text before generating word or character vectors. It performs contraction by simple replacement rules of the commonly used English contractions.

Our comments are in English, so we removed all non-ASCII characters and punctuation. After that, we converted numbers into word shapes and lowercase all the characters.

A dataset of stop words, including 40 terms specific to English sentiment analysis, is used. These words do not have a polarity score, and their removal simplified the text of the comments. As a result, all the stop words in the original dataset have been removed to create a new dataset that does not contain stop words.

In natural language processing, stemming reduces all words to their base form and helps normalize the text. At this step, we stem all the words with the help of the nltk.stem library and replace them in the text of the comment. This change also reduces the number of word vectors.

*B. SMOTE algorithm*

Unbalanced datasets are one of the challenges; the number of one class is higher than the other, which causes the deep learning model to be less accurate on the minority class of the data. This approach is equivalent to the artificial minority oversampling method, done by oversampling the minority class samples or undersampling the majority class samples [18, 42].

SMOTE works by selecting examples close to the feature space, drawing a line between the examples in the feature space, and drawing a new sample at a point along that line. These synthetic training records are generated by randomly selecting one or more of the k-nearest neighbors for each example in the minority class. After the oversampling process, the data is reconstructed, and several classification models can be applied to the processed data.

*C. MCNN for aspect extraction*

This paper used a multi-channel convolutional neural network (MCNN) for the aspect extraction task introduced in [2]. The model has two channels: word embedding and POS embedding channel shown in Fig. 2. For the word embedding, we apply a gensim pre-trained word2vec [43], which is trained using a large corpus of Google news based on the continuous bag of word model architecture. The POS Tag embedding channel's main idea is to better facilitate the sequential tagging process based on [44]. We used a Stanford POS Tagger encoded as a 45- dimensional vector.

After embedding layer, we utilized a convolutional layer that extracts the pivotal features from the user textual reviews. This layer used 128, 5, and 3 filter sizes for the embedding channel and 32, 5, and 3 for the POS channel. We used a dropout rate of 0.5. Next, a pooling operation is applied to capture the maximum elements of the contents. We use softmax and sigmoid as the activation function.

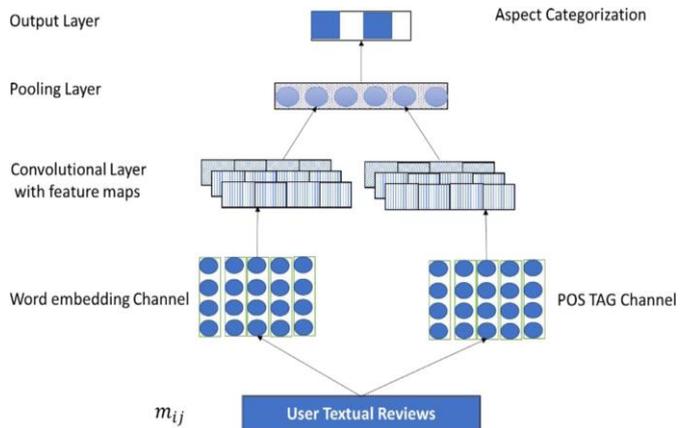

Fig. 2. The framework of MCNN.

*D. Recommender system*

In this paper, we applied tensor decomposition to a weighted aspect rating matrix. To build a recommender system, we used User Based collaborative filtering that automatically recommends items to users based on a decomposed matrix. The whole represented framework is shown in Fig. 3.

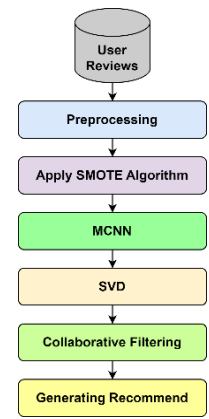

Fig. 3. Our proposed method map.

IV. EVALUATION AND RESULTS

In this section, we demonstrate our dataset and the performance of our proposed method on two levels: our deep neural network and recommender system. Our experiments aim to answer the following questions: (1) what the performance of MCNN is after applying SMOTE algorithm? (2) What is the performance of the proposed model in terms of the rating prediction and Top-5 ranking?

## 1.1. Dataset

We used the Amazon dataset, one of the enormous datasets for Recommender systems. McAuley [45] collected this dataset, comprised of reviews and metadata from 24 individual product categories, and has been used by many researchers. To implement the proposed method, we use three categories of Musical instruments, Automotive, and Instant videos and concatenate them into one.

We removed all records with plain texts or ratings to provide clean datasets. After balancing the data, we randomly categorized our datasets into 80% and 20% for training and testing. The overall ratings of the clean dataset are given in Fig.4.

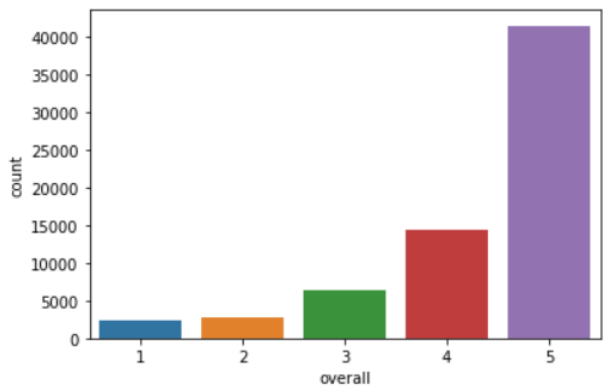

Fig. 4. Distribution of user ratings.

## 1.2. Experiment results

To evaluate our proposed approach's performance in sentiment analysis, we compare our model with the following baselines: (1) CNN + LP [11], the state-of-the-art method using the CNN model for aspect extraction. (2) Pop-dep [6], a method that uses a double propagation approach with word dependency for aspect extraction. (3) MCNN + W2V + POS [2], used pretrained word2vec and POS Tag embedding without applying SMOTE algorithm.

The results of the aspect extraction task have given in Table I. The results are presented in terms of the F1 score, and the best results of each algorithm have been considered.

TABLE I. ASPECT EXTRACTION PERFORMANCE IN TERMS OF F1 SCORE.

| Method | F1 score |
| --- | --- |
| CNN + LP | 90.44% |
| Pop-dep | 88.90% |
| MCNN + W2V + POS | 89.46% |
| MOD 1 | 87% |
| MOD 2 | 84% |
| Proposed method | **91.6%** |

Table I shows that the best performance of the MCNN belongs to the proposed method on the dataset. From Table I, it can be observed that CNN + LP performs better than MCNN without SMOTE algorithm. We also perform MCNN with three parallel channels without using POS tagging (MOD 1) and another without word2vec (MOD 2). MOD 2 performs less accurately than all the other versions in all the cases, and MOD 1 has a lower performance than MCNN with both word2vec and POS tagger. These metrics indicate the impact of using pre-trained word embeddings and POS tag embedding for aspect extraction more than adding a channel to MCNN. These results demonstrate the quality of our extracted aspects for building the recommendation system. Three crucial factors can be considered as the reasons for the better performance of our model: (1) the POS embedding and pre-trained embedding combination helps to facilitate sequence labeling and better capturing of the semantic information of words. (2) SMOTE algorithm helps to balance data and increase the number of training samples, leading the model to better performance.

We use Root Mean Squared Error (RMSE) and Mean Absolute Error (MAE) metrics to evaluate our proposed approach regarding item recommendation. These metrics are given as follows:

$$MAE = \frac{\sum_{(i,j)\in T}|\tilde{R}_{ij}-R_{ij}|}{|T|} \quad (1)$$

$$RMSE = \sqrt{\frac{\sum_{(i,j)\in T}(\tilde{R}_{ij}-R_{ij})^{\Upsilon}}{|T|}} \quad (2)$$

Where T is the testing set, $\tilde{R}_{ij}$ is the predicted rating of item j for user i, and $R_{ij}$ represents the actual value of rating item j for user i. Smaller RMSE / MAE indicates high accuracy.

To evaluate the accuracy of rating prediction, we make a comparison with the following methods: (1) Matrix Factorization [46] (MF), which is widely used as a standard baseline for collaborative filtering. (2) Hidden Factor and Hidden Topics [20] (HFT), which we summarized in section 2. (3) Rating boosted latent topics [47] (RBLT) model, which exploits user opinions rather than weighted preferences from the user textual review. (4) MCNN + word2vector + POS tagging [2], which we mentioned in section 2.

TABLE II. COMPARISON OF THE PROPOSED MODEL RESULTS IN THE RATING PREDICTION.

| Method | RMSE | MAE |
| --- | --- | --- |
| MF | 1.1884 | 0.9631 |
| HFT | 1.0289 | 0.7723 |
| RBLT | 1.0226 | 0.7699 |
| MCNN + W2V + POS | 1.0116 | 0.7577 |
| Proposed method | **0.3291** | **0.1439** |

## V. CONCLUSION

In this paper, to avoid the "cold start" problem, we defined a method using a multi-channel convolutional neural network and the SMOTE algorithm. We have used the proposed opinion mining model based on weighted aspects.

First, we showed how to extract aspect terms using a multi-channel convolutional neural network. Then, we explained how to use the extracted aspects of the user comment text to generate aspect rankings based on the lexicon-based method. In addition, we described how to use aspect ratings to generate weighted opinions and then predict scoring using the SVD method. Using the Amazon dataset, we evaluated the model's feature extraction and product recommendation performance. The evaluation results of the proposed method showed that this model obtained better results compared to the basic techniques in extracting the aspect and recommending the product and reached 91.6% accuracy. The following algorithms can be used to develop a more accurate model.